\documentclass[preprint]{elsarticle}
\DeclareGraphicsExtensions{.pdf,.png}
\usepackage[toc,page]{appendix}
\usepackage{amsmath}
\usepackage{amssymb}
\usepackage{natbib}
\usepackage[english]{babel}
\usepackage{graphicx}
\usepackage{hyperref}
\usepackage{mathtools}

\DeclarePairedDelimiter\floor{\lfloor}{\rfloor}
\usepackage{enumerate}
\DeclareMathOperator*{\minimize}{minimize}
\newcommand{\norm}[1]{\left\lVert#1\right\rVert}
\usepackage{tabularx}
\usepackage{float}
\journal{Arxiv}
\usepackage[section]{placeins}
\newtheorem{thm}{Theorem}
\newtheorem{lem}[thm]{Lemma}
\newtheorem{corollary}[thm]{Corollary}
\newdefinition{rmk}{Remark}
\newdefinition{example}{Example}
\newdefinition{definition}{Definition}
\newproof{pf}{Proof}
\newproof{pot}{Proof of Theorem \ref{thm2}}
\usepackage{color}

\usepackage{hyperref}

\usepackage{subfig}
\usepackage[ruled,vlined]{algorithm2e}
\usepackage{graphics}
\usepackage{wrapfig,booktabs}
\usepackage{tabularx}
\usepackage[utf8]{inputenc}
\usepackage{listings}

\newcommand{\calG}{\mathcal{G}}
\newcommand{\calW}{\mathcal{W}}
\newcommand{\calF}{\mathcal{F}}

\usepackage[hmargin=3cm,vmargin=2cm]{geometry}
\usepackage[utf8]{inputenc}
\usepackage[dvipsnames*,svgnames]{xcolor}
\usepackage{mathtools}
\usepackage{xstring}
\usepackage{xparse}
\usepackage{adjustbox}
\usepackage{varwidth}
\newcounter{tittel}
\newcounter{problem}
\newcounter{alternative}
\newcounter{navn}[problem]

\newif\iffirstalt

\NewDocumentCommand{\Oppgave}{m o}{%
	\IfNoValueTF{#2}{\setcounter{alternative}{0}\stepcounter{problem}    \firstaltfalse}%
	{\stepcounter{alternative}\iffirstalt\else\stepcounter{problem}    \firstalttrue\fi}
	\section*{Oppgave \arabic{problem}%
		{\normalfont\IfNoValueTF{#2}{}{~Alternative     \Roman{alternative}\ }
			\normalsize (#1 poeng)}%
		\addcontentsline{toc}{section}{Oppgave \arabic{problem} }} 
	\vspace{3mm} }

\usepackage{todonotes}

\begin{document}

\begin{frontmatter}

\title{Diagnosis of Small-world Bias in Random Graphs
}

\author{Georgios Argyris}
\address{Department of Science and Technology, University of Twente, Netherlands}

\address{g.argyris@utwente.nl
}

\begin{abstract}
\textbf{Background:} Imagine a paper with $\mathit{n}$ nodes on it where each pair undergoes a coin toss experiment; if heads we connect the pair with an undirected link, while tails maintain the disconnection. This procedure yields a random graph. Now consider duplicating this network onto another paper with a slight bias-a fraction of its links (approximately 1/10) undergo rearrangement. 
If we shuffle the two papers, how can we distinguish the pure random graph from the biased one?
\textbf{Results:}
In response to this challenge, we propose a novel metric called \emph{``Randomness Index" (RI)}. The closer the metric to zero is, the higher degree of randomness in the graph. 
The RI can distinguish between dense small-world networks and dense random graphs; a distinction which is impossible by conventional small-world properties like clustering coefficient and average path length.
To validate its effectiveness, we apply the RI to temporal correlation networks of stock indices. Our findings reveal a reduction in randomness during global economic recession periods.
\textbf{Conclusion:}
The RI emerges as a powerful metric capable of characterizing small-world topology, especially in scenarios where other network measures fail. Beyond its utility in network analysis, the RI is promising for change-point (anomaly) detection in dynamical systems studied by means of multivariate time series.

\end{abstract}

\begin{keyword}
\texttt{Random Graphs, Small-World Networks, Motifs, Graphlets, Graph Isomorphism, Rado Graph, Watts-Strogatz Model, Network Randomness Index, Network Randomness Metric, Multivariate Time Series, Temporal Correlation Networks, Morgan Stanley Capital Index}
\end{keyword}

\end{frontmatter}


\section{Introduction}

\begin{wrapfigure}{r}{0.25\textwidth}
	\hspace{-1.1cm}
	\vspace{-0.8cm}
	\begin{center}
		\includegraphics[scale=.1]{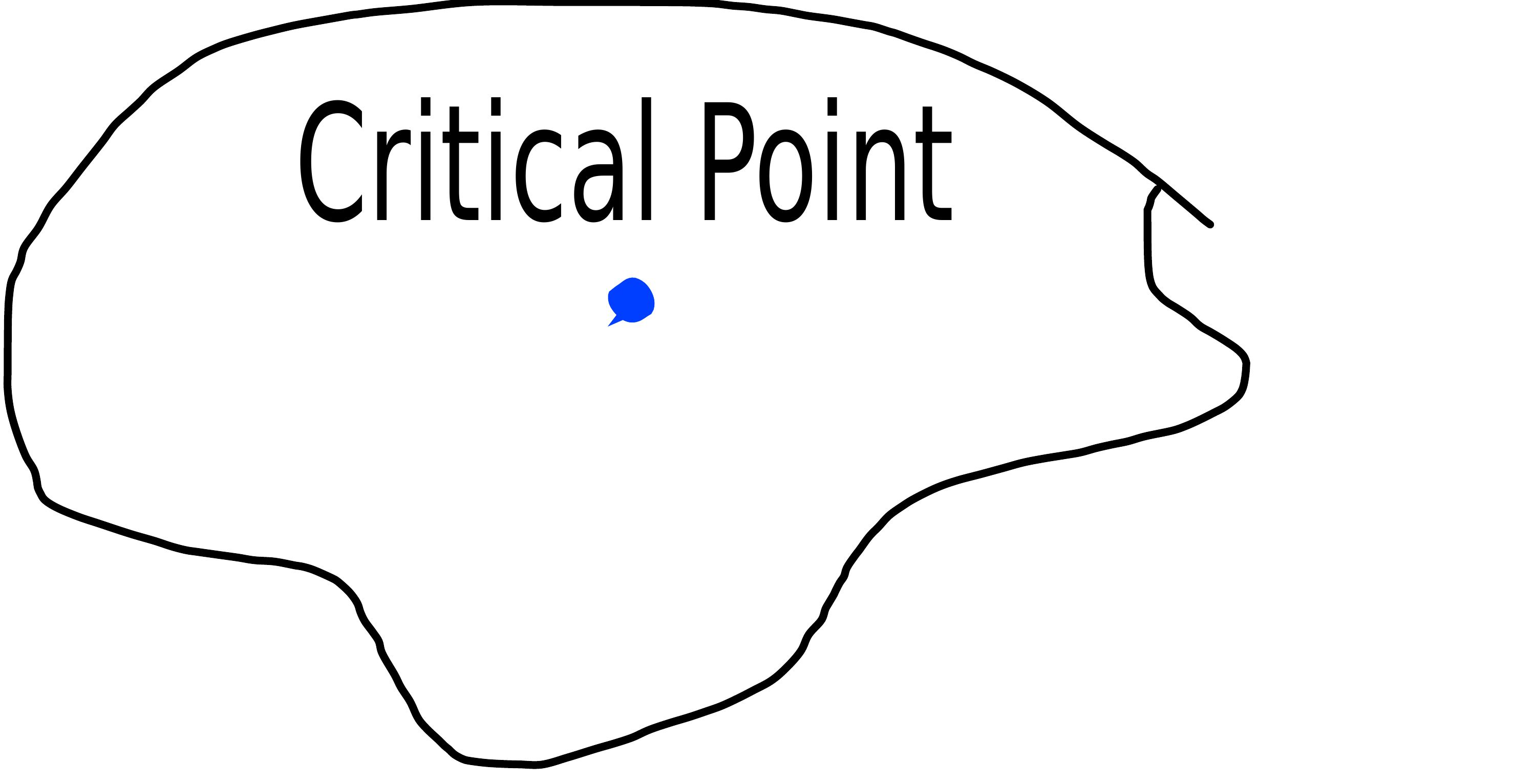}
	\end{center}
	\vspace{-0.4cm}
	\caption{The Critical Point in blue. Whatever is inside the neighbourhood is random, and whatever is outside the neighborhood is small-world.}
	\label{fig:CriticalPoint}
\end{wrapfigure}

\paragraph{Contribution} Small-world topology is characterized by high clustering coefficient and low average path length and constitutes an independent feauture of the inherently-sparse social networks. On the other hand, dense graphs exhibit both high clustering and low average path length (such as the interracial cortical network in the primate brain~\citep{markov2013cortical}) which hinders the discernation of the small-world phenomenon. We propose a method that addresses this problem by distinguishing dense small-world networks from dense random networks through a classification methodology. The latter is based on a critical point and a neighbourhood around this point defined in an euclidean space(see Fig.~\ref{fig:CriticalPoint}). By embedding an observed network into a point of the same euclidean space, we say that an observed network is random if the point lies inside the neighbourhood of the critical point. On the contrary, a network is small-world if it lies outside the neighbourhood. The critical point is a six dimensional point whose coordinates constitute relative frequencies of motifs.

\paragraph{Network types}~\label{sec:preliminaries}
The research of Gilbert, Erdos, and Renyi \citep{Erdos1959, Gilbert1959} on random graph theory inspired the study of real world complex networks. Over the last years, network science provides metrics and measures that characterize either the network or its nodes~\citep{RUBINOV20101059}, and models that describe the different network topologies and generate networks with desirable properties. Without doubt, one of the most popular network topologies is the small-world.
Small-world networks constitute an analogy to the small-world phenomenon of Stanley Milgram~\citep{Milgram} firstly described by the Watts-Strogatz model~\citep{WattsStrogatz1998}. The \emph{Watts and Strogatz (WS) small-world network model} is a three-parameter model $\mathit{(V,k,p_r)}$ where: $\mathit{V}$ is a set of vertices placed on a circular configuration, $\mathit{k}$ denotes the number of neighbors each node is attached, and $\mathit{p_r}$ is the probabilty of rearrangement of a link in the network. On the other hand, the \emph{Erdos-Renyi (ER) random graph model} is considered here as a two-parameter model $\mathit{(V,p)}$: $\mathit{V}$ 
is the number of nodes and $\mathit{p}$ denotes the probability of a link's occurence between any pair of nodes. This article compares the two network topologies in terms of motifs~\cite{motifintro1}.

\begin{wrapfigure}{l}{0.25\textwidth}
	\hspace{-1.1cm}
	\vspace{-0.8cm}
	\begin{center}
		\includegraphics[scale=.1]{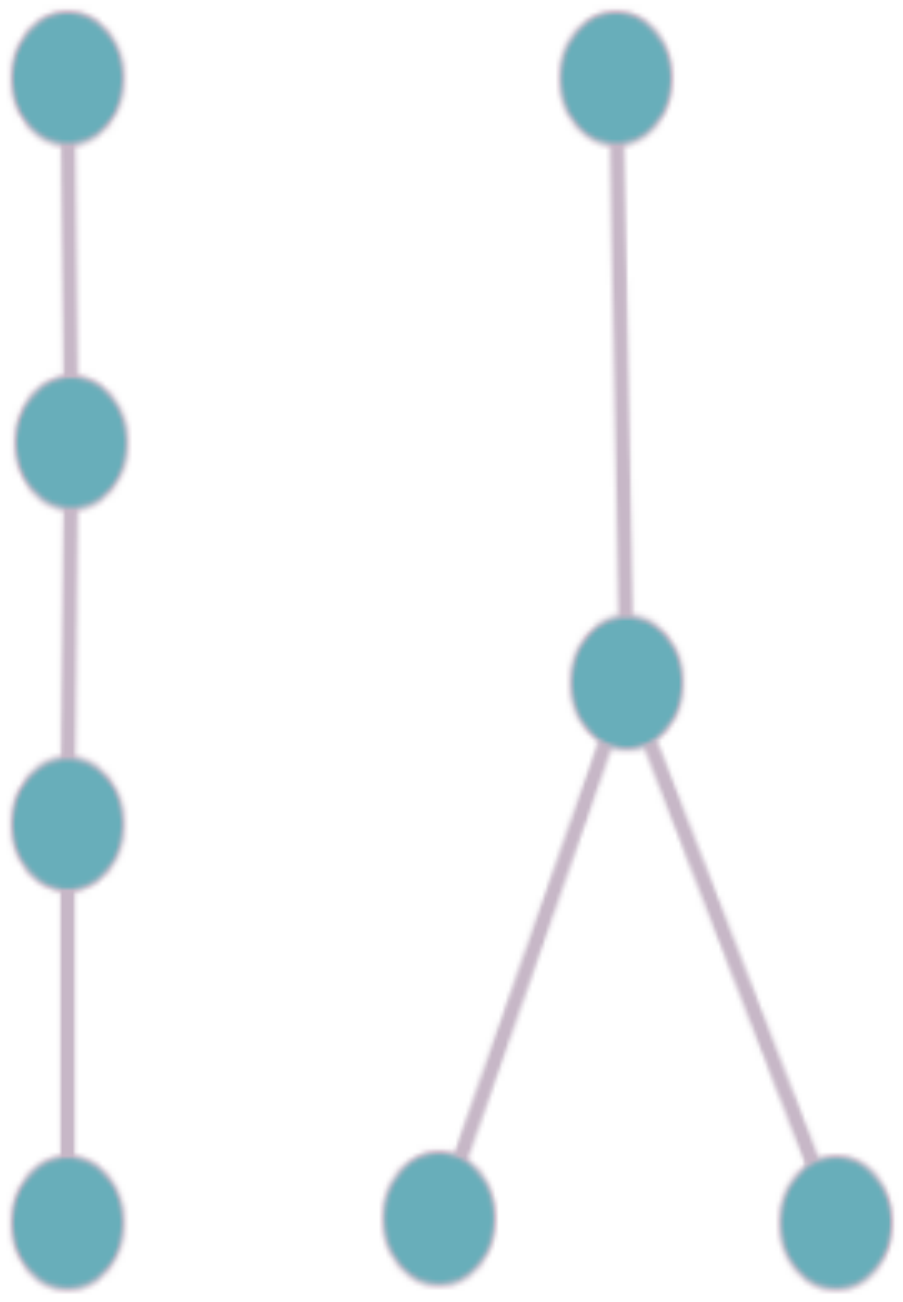}
	\end{center}
	\vspace{-0.9cm}
	\caption{(Left) Sequential pattern. (Right) Star pattern.}
	\label{fig:patterns}
	\vspace{-0.5cm}
\end{wrapfigure}

\begin{figure}[H]
	\centering
	\includegraphics[scale=0.7]{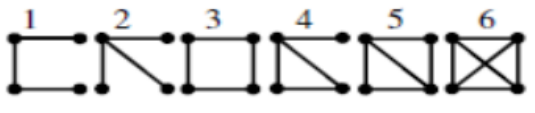}
	\caption{The 6 Motifs; the connected tetrads of nodes}
	\label{fig:motifs}
\end{figure}

\paragraph{Some preliminary thoughts} Consider the two patterns of Fig.~\ref{fig:patterns}, namely the sequential and star pattern on 4 nodes. Which one of these patterns (considered as induced subgraphs) has higher frequency in the pure random graph introduced in the abstract?
To answer this, we need to reconfigure the two motifs into square configurations, like in Fig.~\ref{fig:motifs}, wherein the sequential pattern is Motif 1, and the star pattern is Motif 2. We prove in form of a theorem that, because of symmetries, the sequential pattern has larger frequency than the star pattern.

\paragraph{Motifs and Graphlets}

The patterns shown in Fig.~\ref{fig:motifs} can be found in the case of statistical mechanics as motifs. They were initially explored as \textbf{partial subgraphs} which recur in real-world complex networks with higher frequency than in randomized networks~\citep{motifintro1}. However, in computer science and bioinformatics these patterns are also known as graphlets~\citep{prvzulj2004modeling,prvzulj2007biological} and are considered as small connected non-isomorphic \textbf{induced subgraphs}. 
In this study, we call them \emph{motifs} but \textbf{we strictly refer to induced subgraphs with 4 nodes}. We display the motifs (as in~\citep{small2009transforming}) $m_i$ with $i=1,...,6$ in Fig.~\ref{fig:motifs}.

In the case of bioinfiormatics~\citep{prvzulj2004modeling,prvzulj2007biological}, the authors define relative graphlet frequency distance and compare protein-to-protein interaction networks with various random graph models. In the case of statistical mechanics, a transformation from univariate time series to complex networks has been introduced, and studied the frequency of different connected tetrads of nodes in these networks~\citep{xu2008superfamily}. Motifs' distribution could distinguish different types of continuous dynamics: periodic, chaotic and periodic with noise, as well as chaotic maps, hyperchaotic maps, and noise data when applied to discrete data. Particularly, the authors proved that noisy periodic signals correspond to random visibility networks whereas chaotic time series generate visibility networks that exhibit small-world and scale-free features~\citep{PhysRevLett.96.238701}. 
Inspired by statistical mechanics, our crucial hypothesis is that we can use such patterns to identify the topology of real-world observed networks. Thus, we motivate the connected tetrads of nodes of Fig.~\ref{fig:motifs} (considered as induced subgraphs and called \emph{motifs}) in order to introduce a novel classification method. 

The article is summarized as follows: Section~\ref{sec:NRI}, we introduce the novel Network Randomness Index (NRI)
, in Section~\ref{sec:class}, we validate the efficiency of the NRI by his ability to discern random and small-world networks, in Section~\ref{sec:app} we display how NRI can be used for anomaly detection in multivariate time series, and in Section~\ref{sec: conclusion} we conclude, discuss our findings, and motivate the readers to new horizons for further investigations. All relevant algorithms are implemented in Matlab (see \url{https://github.com/GeorgiosArg/Diagnosis-of-Small-World-Bias-in-Random-Graphs}) and are described in the Appendix.

\section{The Randomness Index}~\label{sec:NRI}

In this section we introduce the Network Randomness Index (NRI) and elaborate on the preliminary thoughts. Particularly, section~\ref{subsec:prelim} provides the background with the preliminary definitions, insection~\ref{subsec:CP} we define the critical point and introduce the NRI 
as the euclidean distance between the critical point and another point that corresponds an observed network, and in Section~\ref{subsec:PT}, we elaborate on the preliminary thoughts by proving that the sequential pattern has larger frequency than the star pattern.

\subsection{Preliminaries}~\label{subsec:prelim}

We start by defining the notion of a graph.

\begin{definition}[Graph]~\label{def:graph}
	A graph $G$ is a pair $(V,E)$ where $V$ is a set of vertices (or nodes), and $E\subseteq V \times V$ is a set of edges (or links), with $|V| \in \mathbb{N}$ being the number of nodes 
	and $|E| \in \mathbb{N}$ being the number of edges.
\end{definition}

We denote with $f_{m_i}$ the number of occurences of the motif $m_i$ considered as induced subgraph in a graph $G$. Next, we define the relative frequency, also introduced in~\citep{orsini2015quantifying} as a measure of randomness, as follows:

\begin{definition}[Relative Frequency]\label{def:RelFr}
	Let $f_{m_i}$ be the frequency of motif $m_i$. We define the \emph{relative frequency} of motif $m_i$ as: 
	\begin{align}
	F_{m_i} = \frac{f_{m_i}}{\sum_{i=1}^{6}{f_{m_i}}}
	\label{eq:2}
	\end{align}
\end{definition}

The critical point is a 6-dimensional point where each coordinate corresponds to the relative frequency of a motif and, for a given graph $G$, is formally defined as follows:

\begin{definition}[Relative Frequency Point (RFP)]\label{def:RFP}
	Let $G$ be a graph and $F_{m_i}$ be the relative frequency of motif ${m_i}$. We define the \emph{relative frequency point} of a graph $G$ as: 
	\begin{align*}
	\calF_G = (F_{m_1}, F_{m_2}, F_{m_3}, F_{m_4}, F_{m_5}, F_{m_6})
	\end{align*}
\end{definition}

We finish this section by reminding the reader that $f_{m_i}$ is the absolute frequency of the motif $m_i$, $F_{m_i}$ its relative frequency ($\mathit{i \in \{1, \ldots, 6\}}$), and $\calF_G$ the RFP of a graph $G$.

\subsection{The critical point}~\label{subsec:CP}

We start by defining the notion of an Erdos-Renyi Random Graph.

\begin{definition}[Erdos-Renyi Random Graph]
	An Erdos-Renyi (ER) random graph (denoted by $\calG$) is a pair $(V,p)$ where $V$ is a set of vertices while $p \in [0,1]$ is the probability of occurence of an edge between any pair of vertices.
\end{definition}

As we have mentioned before, the critical point is a six-dimensional point whose coordinates are the relative frequencies of each motif.  In the case of Erdos-Renyi random graphs, these frequencies can be calculated analytically by closed form formulas. We derive these formulas in this section according to Table~\ref{table:motif_frequencies}. The first column presents the 11 sets of 4-node graphs, where each set is a class of isomorphic graphs. Note that there is a bijection that preserves edges, mapping a pattern to each other pattern within the same set. 
The column \emph{order} contains the number of links of each pattern in class, the \emph{patterns} refers to the number of patterns in each set, in \emph{description} we provide a description of the  motif, and the last column contains the probability of occurence of each pattern. Note that graphs which belong to the same set occur with equal probability in $ER$ graphs, i.e., the probability of occurence of a pattern is a graph invariant under graph isomorphism. 
\begin{table}
\centering
\resizebox{\textwidth}{!}{
\begin{tabular}{ |c|c|c|c|c| } 
 \hline
 \emph{Classes of isomorphic graphs} & \emph{order} & \emph{patterns} & \emph{description} & \emph{probability of pattern occurence} \\
 \hline
 \includegraphics[scale=1.85]{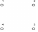} &0& 1& unconnected 
 & $(1-p)^6 $ \\
 \hline
 \includegraphics[scale=0.25]{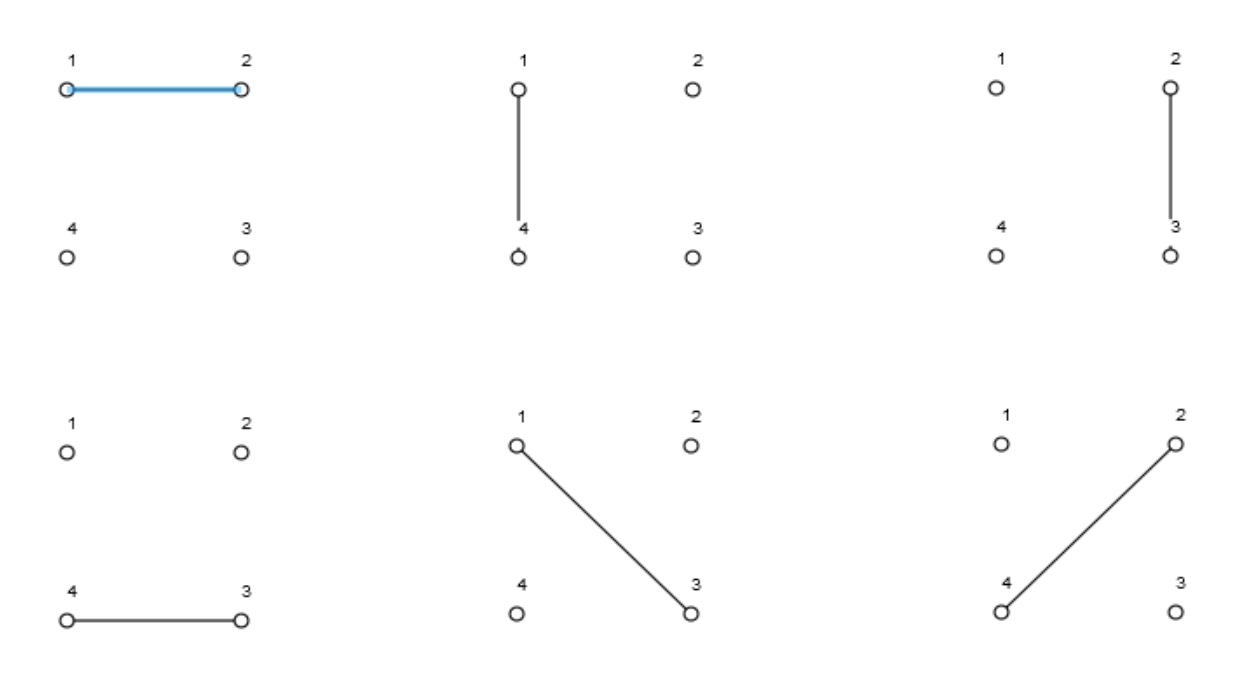} &1& 6& unconnected & $6 \cdot p \cdot (1-p)^5$ \\ 
 \hline
 \includegraphics[scale=0.55]{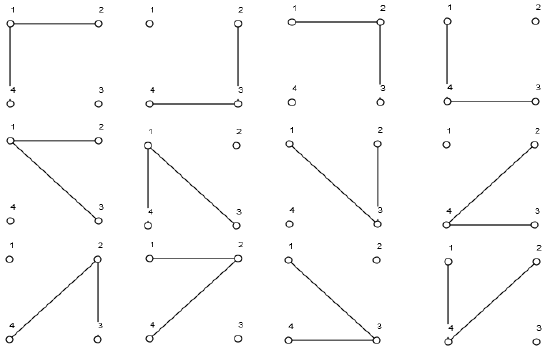} & 2&12& unconnected 
 & $12 \cdot p^2 \cdot (1-p)^4$ \\ 
 \hline
  \includegraphics[scale=0.25]{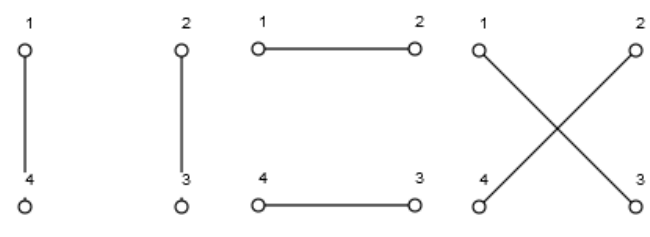} & 2&3& unconnected
  & $3 \cdot p^2 \cdot (1-p)^4$ \\ 
 \hline
  \includegraphics[scale=0.55]{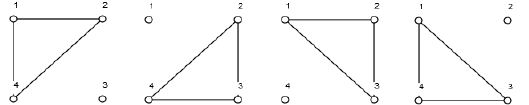} & 3&4& unconnected 
  & $4 \cdot p^3 \cdot (1-p)^3$ \\ 
 \hline
  \includegraphics[scale=0.51]{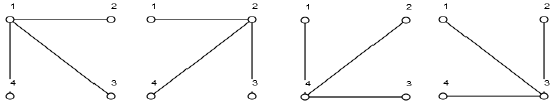} & 3& 4& motif 2; star 
  & $4 \cdot p^3 \cdot (1-p)^3$ \\ 
 \hline
  \includegraphics[scale=0.65]{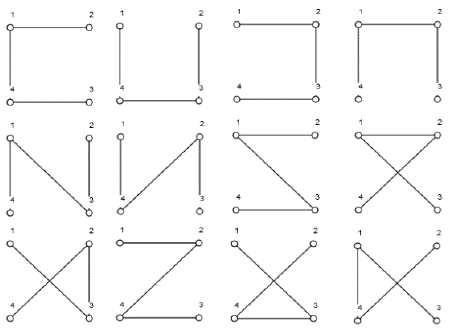} & 3& 12& motif 1; sequentially connected & $12 \cdot p^3 \cdot (1-p)^3$ \\ 
 \hline
  \includegraphics[scale=0.65]{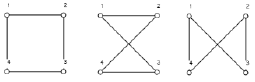} & 4&3& motif 3 
  & $3 \cdot p^4 \cdot (1-p)^2$ \\ 
 \hline
  \includegraphics[scale=0.55]{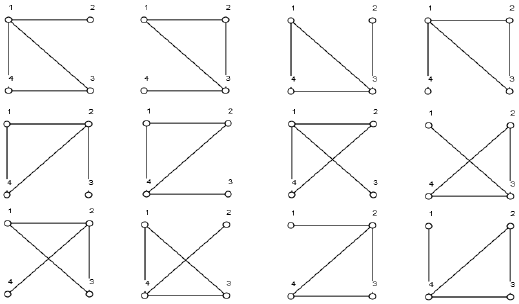} & 4&12& motif 4
  & $12 \cdot p^4 \cdot (1-p)^2$ \\ 
 \hline
  \includegraphics[scale=0.55]{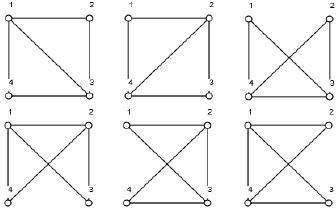} & 5&6& motif 5
  & $6 \cdot p^5 \cdot (1-p)$ \\ 
 \hline
 \includegraphics[scale=1.25]{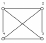} & 6&1& motif 6
 & $p^6$ \\
  \hline
\end{tabular}}
\caption{Detailed description of all 4-node patterns.}
\label{table:motif_frequencies}
\end{table}
For a network of size $n$, there are \(\binom{n}{4} \) tetrads of nodes. If we denote with \( N_i\) the number of isomorphs the motif $i$ has, and \( l_i\) its order (the number of links of $i$), the absolute frequency \( f_{m_i}\) is given by the following lemma.

\begin{lem}\label{lem:RF}
	 Let $\mathcal{G}=(V,p)$ be an Erdos-Renyi random graph with $|V|=n >4$ and $m_i=(4,l_i)$ be a motif (induced subgraph) on $4$ nodes with $l_i$ edges. We denote with $N_{m_i}$ denote the number of isomorphic patterns of motif $m_i$. The expected frequency of motif $m_i$ in the random graph $\calG$ is given by the following formula:
	\begin{equation}\label{eq:frequency}
	f_{m_i}(n,p) = \binom{n}{4}*N_{m_i}*p^{l_i}*(1-p)^{6-l_i}
	\end{equation}
\end{lem}

A formula akin to equation~\ref{eq:frequency} was previously established for directed networks in~\citep{itzkovitz2003subgraphs}. In this study, the authors, employing approximations for the average number of subgraphs in an ensemble of random networks with an arbitrary degree sequence, reached the conclusion that specific subgraphs tend to occur more frequently in real-world and scale-free networks featuring a specified power-law node degree distribution compared to randomized graphs. Moreover, they consider the number of isomorphic patterns $\mathit{N_{m_i}}$ as a parameter $\mathit{\lambda}$ of order $\mathit{1}$ $(\mathit{O(1)})$.
Nevertheless, these values merely represent the number of isomorphic patterns that a motif (induced subgraph) can exhibit.


\textbf{We denote with $\calF_{\calG}(n,p)$ the RFP of the Erdos-Renyi graph and call it critical point} which is the point of Fig.~\ref{fig:CriticalPoint}. The uniqueness of the critical point is secured by the following theorem of Erdos and Rado.

\begin{thm}[Rado Graph]~\label{thm:Rado}
	Let $\calG=(V,p)$ be the random graph over a countably infinite set of vertices ($|V|=n=\infty$) 
	for some fixed $p$. The Rado graph $\calG_\infty=(\infty,p)$ 
	is unique under isomorphism.
\end{thm}

In words, the theorem guarantees the following counter-intuitive phenomenon; if several people have an infinitely big paper with infinitely many nodes on it and, for each pair of nodes, they perform the toin coss experiment mentioned in the abstract, then everybody will end up with the same graph (under isomorphism). The previous theorem can be found in~\cite{editionreinhard} (pages 228-229). The following Corollary immediately follows:

\begin{corollary}[Critical Point]
	Let $\calG_\infty$ be the Rado graph.
	The $\calF_{\calG_\infty}$ is unique.
\end{corollary}

The following remark is crucial for the development of the classification method of Section~\ref{sec:class}.

\begin{rmk}[Network Randomness Index]~\label{Remark:NRI}
	Consider an observed real world network $G=(V,E)$ and $\calF_G$ its RFP. The greater the proximity of the observed network's frequency vector, denoted as $\calF_G$, to the critical point $\calF_{\calG}$ in terms of Euclidean distance ($\norm{\calF_{\calG}-\calF_G}$), the higher the randomness exhibited by the observed network. This happens because for a random graph on a finite $n$ number of vertices, the point $\calF_{\calG}$ lies in a concrete neighborhood (see Fig.~\ref{fig:CriticalPoint}).
\end{rmk}

\subsection{On the preliminary thoughts}~\label{subsec:PT}

We next provide a theorem that relates the frequencies of subgraphs in the Rado graph, with their isomorphs. The following theorem gives also the solution to the problem stated in the preliminary thoughts of the introduction.

\begin{thm}
	Let $G_1=(V_1,E_1)$, $G_2=(V_1,E_1)$ be two graphs with the same number of nodes $|V_1|=|V_2|=m$ and the same number of edges $|E_1|=|E_2|=l$. The expected frequency of $G_1$ is higher than the expected frequency of $G_2$ in a random graph $\calG=(V,p)$ with $|V|=n \geq m$ if and only if $G_1$ has more isomorpfic patterns than $G_2$, i.e., $N_{G_1} \geq N_{G_2}$.
\end{thm}

\begin{pf}
	We denote with $N_{G_1}$, $N_{G_2}$ the number isomorphic patterns, and $f_{G_1}$, $f_{G_2}$ the relative frequency of the graphs $G_1$, $G_2$ respectively. We also denote with $m$ the number of nodes which is equal. 
	It holds that:
	\begin{gather*}
	f_{G_1}<f_{G_2}
	\Leftrightarrow \\
	\Leftrightarrow N_{G_1} \cdot \binom{n}{m} \cdot p^l \cdot (1-p)^{\frac{m(m-1)}{2}-l}
	<N_{G_2} \cdot \binom{n}{m} \cdot p^l \cdot (1-p)^{\frac{m(m-1)}{2}-l} 
	\Leftrightarrow \\
	\Leftrightarrow N_{G_1}<N_{G_2}
	\end{gather*}
\end{pf}

The following remark immediately follows.
\begin{rmk}
    Considering the preliminary thoughts of the Section~\ref{sec:preliminaries}, it is straightforward that the sequential motif has higher frequency than a star motif in the Rado graph. We refer the reader to Table~\ref{table:motif_frequencies} in the column \emph{patterns}. Notice that the star motif has $4$ patterns while the sequential motif has $12$ patterns.
\end{rmk}

Density is a critical parameter that can dramatically change the frequency of motifs. It is particularly easy to calculate.

\begin{definition}[Density]~\label{def:density}
	Let $G=(V,E)$ be a subgraph with $|V|=m$ and $|E|=l$. The density of $G$ is defined as follows:
	\[
	d_G = \frac{l}{\frac{m * (m-1)}{2}} 
	\]
\end{definition}

Next, we provide a Theorem that guarantees that a subgraph achieves its maximum frequency in a random graph when the density of the subgraph equals with the parameter $p$ of the random graph i.e. the probability of occurrence of an edge. To prove this we use the maximum likelihood estimation.

\begin{thm}[Maximum Frequency of a Graph]
	Let $G=(U,E)$ be a graph with $N_G$ isomorphic patterns $|U|=m$, $|E|=l$, and $\calG=(V,p)$ be a random graph with $|V|=n > m$. The finite induced graph $G$ maximizes its frequency in $\calG$ when $d_G = p$.
\end{thm}

\begin{pf}
	The frequency of $G$ in the random graph $\calG$ is given by the following formula:
	\[
	F_G = N_G * \binom{n}{m} * p^l * (1-p)^{\frac{m*(m-1)}{2}-l}
	\]
	The function $log(x)$ is monotonic on $x$ and, hence, it holds that $max (F_G) = max (log(F_G))$. We first calculate $log(F_G)$:
	\begin{align*}
		log(F_G) &= log(N_G * \binom{n}{m} * p^l*(1-p)^{\frac{m*(m-1)}{2}-l}) \iff\\
		& = log(N_G) + log(\binom{n}{m}) + log(p^l) + log((1-p)^{\frac{m*(m-1)}{2}-l})\iff\\
		& = log(N_G) + log(\binom{n}{m}) + l*log(p) + (\frac{m*(m-1)}{2}-l) * log(1-p)
	\end{align*}
	To find the maximum in terms of the probability $p$, we calculate the derivative of $log(F_G)$ up to $p$ and equalize it to zero:
	\begin{align*}
		\frac{\partial log(F_G)}{\partial p} = 0 \implies
	\end{align*}
	\begin{align*}
		l*\frac{1}{p} + (\frac{m*(m-1)}{2}-l) * \frac{1}{1-p} = 0 \implies (p-1) * l + p * (\frac{m*(m-1)}{2}-l)=0 \implies \\
		p * l - l + p * (\frac{m*(m-1)}{2}-l)=0 \implies p * l + p * (\frac{m*(m-1)}{2}-l) = l \implies \\ 
		p * (l+\frac{m*(m-1)}{2}-l) = l \implies p * (\frac{m*(m-1)}{2}) = l \implies p  = \frac{l}{\frac{m*(m-1)}{2}}
	\end{align*}
	Therefore, we conclude that the frequency is maximized when $p=d_G$ (see Definition~\ref{def:density}).
\end{pf}

\section{Classification of Small-World and Random Graphs}~\label{sec:class}


In this section, we present a novel classification method designed to differentiate between small-world and random networks. Section~\ref{sec:sw} introduces the Watts-Strogatz small-world networks along with the classification method. The validation of our method through Monte Carlo simulations is carried out in Section~\ref{sec:experiments}. Further refinement of the Monte Carlo experimental validation, aimed at assessing the classification power of our framework, is discussed in Section~\ref{sec:ref}.

\subsection{Small-world Networks and Classification Methodology}~\label{sec:sw}

\paragraph{Small-world Networks} We first give the definition of a small-world network.

\begin{figure}[H]
	\centering
	\includegraphics[scale=0.5]{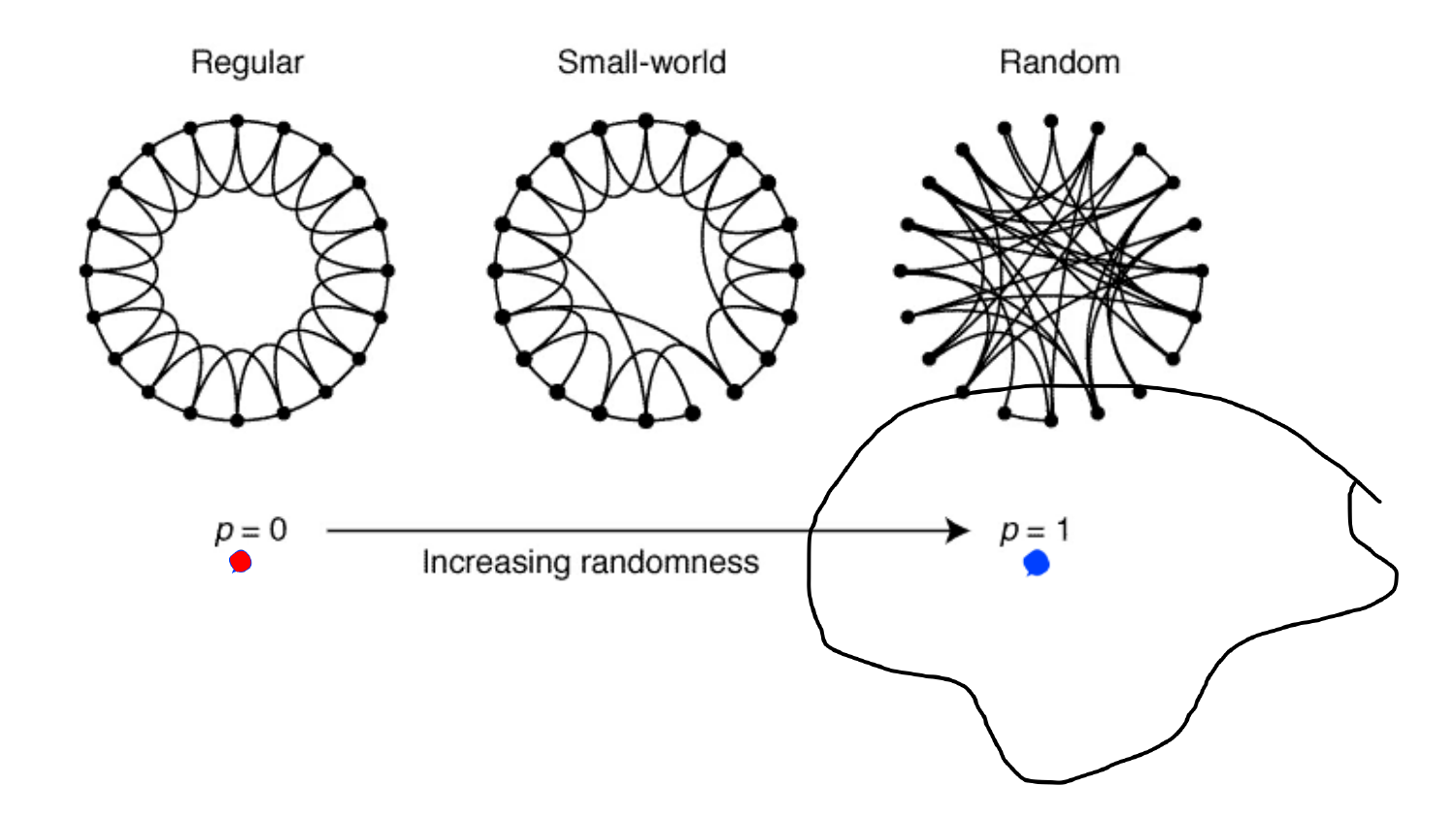}
	\vspace{-0.3cm}
	\caption{Visual representation of the relative frequency point of the regular lattice to the critical point. The top part of the figure was takes from~\cite{WattsStrogatz1998}. Note that $p_r$ is denoted by $p$ in the figure.}
	\label{fig:cp2}
\end{figure}

\begin{definition}[Watt-Strogatz Small-world Networks]~\label{def:swgraph}
A small-world network, denoted by $\calW$, is a triple $(V,k,p_r)$ where $V$ is a set of nodes circular configuration (see top part of Fig.~\ref{fig:cp2}), $k$ is the number of neighbors that each node is attached, and $p_r \in [0,1]$ the probability of rearrangement of an edge.
\end{definition} 

Consider the \emph{regular} lattice depicted in the upper left part of Fig.~\ref{fig:cp2}. This lattice forms a small-world network denoted as $\calW$ with $\mathit{k=2}$-each node is connected to precisely two of its nearest neighbors. Additionally, $\mathit{p_r=0}$ since no edges undergo rearrangement. Moving to the upper middle part of the figure, we see a small-world network with $\mathit{p_r=0.1}$. The top right part of the figure illustrates a $\calW$ with $\mathit{p_r=1}$.

Small-world networks, characterized by different values of $\mathit{p_r}$, exhibit unique relative frequency points in the six-dimensional space of motif frequencies. Specifically, when $\mathit{p_r=0}$, the corresponding Relative Frequency Point (RFP) lies outside the neighborhood of the Critical Point depicted in Fig.\ref{fig:CriticalPoint} (highlighted by the red point in the bottom left part of Fig.\ref{fig:cp2}). However, as $\mathit{p_r}$ increases, the RFP progressively approaches the neighborhood of the critical point. In the limit where $p_r \rightarrow 1$, it crosses the boundary and enters the neighborhood, as indicated by the blue point in Fig.~\ref{fig:cp2}.



\textbf{The Relative Frequency Point (RFP) of a small-world network is denoted as $\calF_{\calW}$}, representing a function of the three parameters ($\mathit{n, p_r, k}$), as defined in Definition~\ref{def:swgraph}. The empirical computation of $\calF_{\calW}$ involves generating $\mathit{100}$ networks with varying values of $\mathit{p_r}$ and $\mathit{k}$ for a standard $\mathit{n}$. The RFP is then calculated for each of these networks, and $\calF_{\calW}(n, p_r, k)$ is determined as the average across this set of $\mathit{100}$ networks. The detailed steps for this empirical computation are elucidated in Algorithm~\ref{alg:RFP_SW} in Appendix~\ref{appendix:A}.

\paragraph{Hypothesis} The crucial hypothesis of this section is the following: \emph{suppose that we have an observed real-world network, could we find which model generates (approximates) it?} To answer this question we develop the following classification methodology. 

\paragraph{The Classification Methodology}~\label{sec:classification}
Considering an observed network as a graph $\mathit{G=(V,E)}$, as defined in Def.~\ref{def:graph}, our approach for identifying the model that best approximates an observed network is summarized in the following $\mathit{2}$ steps.

\begin{enumerate}[Step 1:]
\item 
To begin, we calculate the RFP of the observed network, denoted as $\calF_G$, utilizing the \emph{ComputeRFP} routine (see~\ref{routine: ComputeRFP}). Then, we apply the Embedding Algorithm, detailed in Appendix~\ref{sec:RFP_top}, to map the network topologies into their corresponding RFPs for the standard size $\mathit{n=|V|}$ of the observed network. For instance, the random graph is positioned within the critical point $\calF_{\calG}$, whereas the small-world network finds its place at another point, denoted by $\calF_{\calW}$.
\item Following the insights of Remark~\ref{Remark:NRI}, the generative model (target function) of the observed network is the one that minimizes the euclidean distance between the RFP $\calF_G$ of step 1 and the RFP of the corresponding model. In other words, the observed network is random if $\norm{\calF_{\calG}-\calF_G} < \norm{\calF_{\calW}-\calF_G}$ and small-world if $\norm{\calF_{\calG}-\calF_G} > \norm{\calF_{\calW}-\calF_G}$.
\end{enumerate}

We propose that the previous classification method of an observed network to its approximated topology (small-world or random). The method is summarized in Algorithm~\ref{alg:classification_method}. 

\subsection{Experimental Validation with Monte Carlo Simulations}~\label{sec:experiments}


In this section, we validate our classification method through Monte Carlo simulations. This involves (i) generating a substantial number of networks using each model, with variation in each model parameter (refer to the explanation in \emph{Configuration}), (ii) presenting the \emph{Results}, which detail the models to which the generated networks are mapped using our classification method, and (iii) drawing conclusions from these results in the \emph{Interpretation}.

\paragraph{Configuration} We generate networks for each one of the models ER $(n,p)$, WS $(n,k,p_r)$ and for each one of their parameters $n,p, k,p_r$. Particularly, for $\mathit{n=25}$, $\mathit{50}$ we generate $\mathit{100}$ networks but for $\mathit{n=75}$ we generate only $\mathit{50}$ networks for better computational efficiency. We validate our method to $\mathit{56.750}$ generated networks overall: 
\begin{itemize}
	\item $\mathit{12.900}$ networks of size $\mathit{n=25}$ ($\mathit{900}$ ER, i.e., $\mathit{100}$ for each $\mathit{p=0.1, \ldots, 0.9}$ + $\mathit{12.000}$ WS, $\mathit{100}$ for each $\mathit{k=2, \ldots, 24}$ for each $\mathit{p_r=0, 0.1, \ldots, 0.9}$),
	\item $\mathit{24.900}$ networks of size $\mathit{n=50}$ ($\mathit{900}$ ER, i.e., $\mathit{100}$ for each $\mathit{p=0.1, \ldots, 0.9}$ + $\mathit{24.000}$ WS, $\mathit{100}$ for each $\mathit{k=2, \ldots, 48}$ for each $\mathit{p_r=0, 0.1, \ldots, 0.9}$),
	\item $\mathit{18.950}$ networks of size $\mathit{n=75}$, ($\mathit{900}$ ER, i.e., $\mathit{100}$ for each $\mathit{p=0.1, \ldots, 0.9}$ + $\mathit{18.500}$ WS, $\mathit{50}$ for each $\mathit{k=2, \ldots, 74}$ for each $\mathit{p_r=0, 0.1, \ldots, 0.9}$).
\end{itemize}
We then classify each one of these networks according to the classification methodology~\ref{sec:classification}~\footnote{For the experimental validation, we also considered scale-free networks constructed both by the Barabasi-Albert preferential attachment mechanism and another method that constructs scale-free networks with exponential degree distribution. However, we do not present these results because it is out of the scope of this article. The full results can be found on~\url{https://github.com/GeorgiosArg/Diagnosis-of-Small-World-Bias-in-Random-Graphs}. Scale-free networks can be identified by naked eye because of their low density.
}.

\paragraph{Results}

We present part of the results in the following tables. 
In the first row, the initial cell contains the size of the generated networks ($\mathit{n}$), while the remaining cells specify the models employed for generating the $\mathit{100}$ networks. The models to which the classification method maps the generated networks are presented in the first column. Specifically, \emph{ER} corresponds to Erdos-Renyi, with the adjacent number indicating the probability ($\mathit{p}$) of edge occurrence. \emph{WS} represents Watts-Strogatz, where the first number adjacent to it is the probability of edge rearrangement ($\mathit{p_r}$), and the second number denotes the value of parameter $\mathit{k}$-the number of neighbors to which each node is attached.


\begin{table}[H]
	\centering
	\scalebox{0.8}{
		\begin{tabular}
			{l|ccccccccc}
			\hline
			$n=25$ &    ER 0.1    & ER 0.2 & ER 0.3 & ER 0.4 & \textbf{ER 0.5} & ER 0.6 & ER 0.7 & ER 0.8 & ER 0.9 \\ \hline
			ER 0.1    & 3      & 1      & 0      & 0      & 0      & 0      & 0      & 0      & 0  \\ \hline
			ER 0.2    & 5      & 31     & 2      & 0      & 0      & 0      & 0      & 0      & 0  \\ \hline
			ER 0.3    & 0      & 7      & 56     & 4      & 0      & 0      & 0      & 0      & 0  \\ \hline
			ER 0.4    & 0      & 0      & 2      & 48     & 0      & 0      & 0      & 0      & 0  \\ \hline
			\textbf{ER 0.5}    & 0      & 0      & 0      & 0      & \textbf{50}     & 1      & 0      & 0      & 0  \\ \hline
			ER 0.6    & 0      & 0      & 0      & 0      & \textbf{1}      & 61     & 0      & 0      & 0  \\ \hline
			ER 0.7    & 0      & 0      & 0      & 0      & 0      & 1      & 50     & 2      & 0  \\ \hline
			ER 0.8    & 0      & 0      & 0      & 0      & 0      & 0      & 1      & 66     & 0  \\ \hline
			ER 0.9    & 0      & 0      & 0      & 0      & 0      & 0      & 0      & 1      & 51 \\ \hline
			WS 0.8,3  & 0      & 8      & 8      & 0      & 0      & 0      & 0      & 0      & 0  \\ \hline
			WS 0.8,4  & 0      & 0      & 6      & 1      & 0      & 0      & 0      & 0      & 0  \\ \hline
			WS 0.8,5  & 0      & 0      & 2      & 14     & 0      & 0      & 0      & 0      & 0  \\ \hline
			\textbf{WS 0.8,6}  & 0      & 0      & 0      & 4      & \textbf{12}     & 0      & 0      & 0      & 0  \\ \hline
			WS 0.9,2  & 5      & 4      & 0      & 0      & 0      & 0      & 0      & 0      & 0  \\ \hline
			WS 0.9,3  & 0      & 6      & 2      & 0      & 0      & 0      & 0      & 0      & 0  \\ \hline
			WS 0.9,4  & 0      & 0      & 4      & 2      & 0      & 0      & 0      & 0      & 0  \\ \hline
			WS 0.9,5  & 0      & 0      & 1      & 16     & 0      & 0      & 0      & 0      & 0  \\ \hline
			\textbf{WS 0.9,6}  & 0      & 0      & 0      & 2      & \textbf{22}     & 0      & 0      & 0      & 0  \\ \hline
			WS 0.9,7  & 0      & 0      & 0      & 0      & \textbf{4}      & 19     & 0      & 0      & 0  \\ \hline
			WS 0.9,8  & 0      & 0      & 0      & 0      & 0      & 3      & 17     & 0      & 0  \\ \hline
			WS 0.9,9  & 0      & 0      & 0      & 0      & 0      & 0      & 7      & 11     & 0  \\ \hline
			WS 0.9,10 & 0      & 0      & 0      & 0      & 0      & 0      & 0      & 12     & 9  \\ \hline
			WS 0.9,11 & 0      & 0      & 0      & 0      & 0      & 0      & 0      & 0      & 38 \\ \hline
			Overall   & 13     & 57     & 83     & 91     & \textbf{89}     & 85     & 75     & 92     & 98 \\ \hline
	\end{tabular}}
	\caption{Classification of $\mathit{900}$ \emph{ER} networks with $\mathit{n=25}$ nodes; $\mathit{100}$ networks as $\mathit{p}$ varies ($\mathit{0.1, 0.2 \ldots 0.9}$).}
	\label{table:table2}
\end{table}

It is evident that when $\mathit{p=0.1}$, the method struggles to classify effectively due to the absence of formed motifs. As $\mathit{p}$ increases, facilitating motif formation, the method achieves more accurate classification for nearly $\mathit{50\%}$ of the networks. Notice, however, that most of the networks that are not classified to the correct model, are classified to ``relative" models, i.e., WS models with high probability of rearrangement of an edge, and similar density. For example, if $\mathit{p=0.5}$, the generated network has expected density $\mathit{d_G=0.5}$. A WS network has density $\mathit{d_{\calW}=\frac{n \times k}{\frac{n\times(n-1)}{2}}}$ and, hence, if $\mathit{d_{\calW}=0.5}$, we have that $\mathit{k=6}$. We display the relevant situation with bold in the corresponding entries of Table~\ref{table:table2}.

For $\mathit{n=50}$, we observe the same phenomenon that we present in Table~\ref{table:table6}.
For example, considering the case of $\mathit{100}$ ER $\mathit{(25,0.5)}$ graphs, we notice that $\mathit{58}$ networks are correctly classified to their generative model, but $\mathit{41}$ networks are classified to the related models of similar equal density.

\paragraph{Interpretation} The classification method fails to classify correctly a large amount of ER graphs because ER graphs are mapped to ``related" models.
This phenomenon arises due to the combinatorial explosion of generative models. To mitigate the combinatorial explosion, we decrease the number of possible generative models by considering known the density of the generated networks.

\begin{table}[H]
	\centering
	\scalebox{0.8}{%
		\begin{tabular}{l|cccccccc}
			\hline
			$n=50$ &   ER 0.1     & ER 0.2 & ER 0.3 & ER 0.4 & \textbf{ER 0.5} & ER 0.6 & ER 0.7 & ER 0.8  \\ \hline
			ER 0.1     & 47     & 0      & 0      & 0      & 0      & 0      & 0      & 0  \\ \hline
			ER 0.2     & 0      & 40     & 0      & 0      & 0      & 0      & 0      & 0  \\ \hline
			ER 0.3     & 0      & 0      & 59     & 0      & 0      & 0      & 0      & 0  \\ \hline
			ER 0.4     & 0      & 0      & 0      & 58     & 0      & 0      & 0      & 0  \\ \hline
			\textbf{ER 0.5}     & 0      & 0      & 0      & 0      & \textbf{58}     & 0      & 0      & 0  \\ \hline
			ER 0.6     & 0      & 0      & 0      & 0      & 0      & 49     & 0      & 0  \\ \hline
			ER 0.7     & 0      & 0      & 0      & 0      & 0      & 0      & 61     & 0  \\ \hline
			ER 0.8     & 0      & 0      & 0      & 0      & 0      & 0      & 0      & 67 \\ \hline
			WS 0.9, 2  & 4      & 0      & 0      & 0      & 0      & 0      & 0      & 0  \\ \hline
			WS 0.9, 3  & 9      & 0      & 0      & 0      & 0      & 0      & 0      & 0  \\ \hline
			WS 0.9, 4  & 11     & 0      & 0      & 0      & 0      & 0      & 0      & 0  \\ \hline
			WS 0.9, 5  & 0      & 34     & 0      & 0      & 0      & 0      & 0      & 0  \\ \hline
			WS 0.9, 6  & 0      & 12     & 0      & 0      & 0      & 0      & 0      & 0  \\ \hline
			WS 0.9, 7  & 0      & 0      & 13     & 0      & 0      & 0      & 0      & 0  \\ \hline
			WS 0.9, 8  & 0      & 0      & 21     & 0      & 0      & 0      & 0      & 0  \\ \hline
			WS 0.9, 9  & 0      & 0      & 0      & 7      & 0      & 0      & 0      & 0  \\ \hline
			WS 0.9, 10 & 0      & 0      & 0      & 20     & 0      & 0      & 0      & 0  \\ \hline
			WS 0.9, 11 & 0      & 0      & 0      & 12     & 0      & 0      & 0      & 0  \\ \hline
			\textbf{WS 0.9, 12} & 0      & 0      & 0      & 0      & \textbf{26}     & 0      & 0      & 0  \\ \hline
			WS 0.9, 13 & 0      & 0      & 0      & 0      & \textbf{15}     & 0      & 0      & 0  \\ \hline
			WS 0.9, 14 & 0      & 0      & 0      & 0      & 0      & 16     & 0      & 0  \\ \hline
			WS 0.9, 15 & 0      & 0      & 0      & 0      & 0      & 33     & 0      & 0  \\ \hline
			WS 0.9, 16 & 0      & 0      & 0      & 0      & 0      & 0      & 1      & 0  \\ \hline
			WS 0.9, 17 & 0      & 0      & 0      & 0      & 0      & 0      & 15     & 0  \\ \hline
			WS 0.9, 18 & 0      & 0      & 0      & 0      & 0      & 0      & 5      & 0  \\ \hline
			WS 0.9, 19 & 0      & 0      & 0      & 0      & 0      & 0      & 0      & 10 \\ \hline
			WS 0.9, 20 & 0      & 0      & 0      & 0      & 0      & 0      & 0      & 7  \\ \hline
			Overall    & 71     & 86     & 93     & 97     & \textbf{99}     & 98     & 82     & 84 \\ \hline
		\end{tabular}
	}
	\caption{Classification of $\mathit{900}$ \emph{ER} networks with $\mathit{n=50}$ nodes; $\mathit{100}$ networks as $\mathit{p}$ varies ($\mathit{0.1, 0.2 \ldots 0.9}$).}
	\label{table:table6}
\end{table}

\subsection{Refinement of the Experimental Validation}\label{sec:ref}

Considering the density of an observed network known, we reduce the number of possible generative models by calculating the parameter $\mathit{p}$ of the Erdos-Renyi random graph, and the parameter $\mathit{k}$ of the Watts-Strogatz model. Let $\mathit{d_G}$ denote the density of an observed network.


\paragraph{Configuration} The configuration of the refined experimental validation is the same as in~\ref{sec:experiments}; we generate the same amount of networks ($\mathit{56.750}$). However, we first compute their density ($\mathit{d_G}$) before we classify them with our method. With this, we can calculate the parameters $\mathit{p}$ of the ER graphs and $\mathit{k}$ of the WS graphs; If we want to approximate the density of the observed network with the ER model we have to set $\mathit{p=d_{\calG}=d_G}$. If we want to approximate the density of the observed network with the WS model, we have to set:
\begin{align*}
& d_{\calW}=d_G \implies
d_{G}=\frac{n \times k}{\frac{n\times(n-1)}{2}} \implies\\
& k=\frac{d_G \times (n-1)}{2}
\end{align*}

\paragraph{Results} 
The results are presented as in previous section; in the first row, the initial cell contains the size of the generated networks ($\mathit{n}$), while the remaining cells specify the models employed for generating the $\mathit{100}$ networks. The models to which the classification method maps the generated networks are presented in the first column.

\begin{table}[H]
	\resizebox{\textwidth}{!}{%
		\begin{tabular}{llllllllll}
			\hline
		n=25	&   ER 0.1 & ER 0.2 & ER 0.3 & ER 0.4 & ER 0.5 & ER 0.6 & ER 0.7 & ER 0.8 & ER 0.9  \\ \hline
			ER     & 55     & 64     & 72     & 79     & 88     & 90     & 97     & 97     & 96 \\ \hline
			WS 0.6 & 11     & 3      & 3      & 3      & 1      & 2      & 0      & 0      & 0  \\ \hline
			WS 0.7 & 4      & 1      & 7      & 0      & 1      & 2      & 0      & 0      & 0  \\ \hline
			WS 0.8 & 1      & 3      & 3      & 16     & 10     & 6      & 3      & 3      & 4  \\ \hline
			WS 0.9 & 12     & 21     & 11     & 0      & 0      & 0      & 0      & 0      & 0  \\ \hline
		\end{tabular}%
	}	\caption{Classification of $\mathit{900}$ \emph{ER} networks with $\mathit{n=25}$ nodes; $\mathit{100}$ networks as $\mathit{p}$ varies ($\mathit{0.1, 0.2 \ldots 0.9}$).}
 \label{Table4}
\end{table}

\begin{table}[H]
	\resizebox{\textwidth}{!}{%
		\begin{tabular}{llllllllll}
			\hline
		n=50	&  ER 0.1 & ER 0.2 & ER 0.3 & ER 0.4 & ER 0.5 & ER 0.6 & ER 0.7 & ER 0.8 & ER 0.9   \\ \hline
			ER     & 74     & 75     & 83     & 85     & 96     & 97     & 95     & 99     & 95 \\ \hline
			WS 0.6 & 5      & 2      & 0      & 0      & 0      & 0      & 0      & 0      & 1  \\ \hline
			WS 0.7 & 3      & 4      & 1      & 0      & 0      & 0      & 0      & 0      & 1  \\ \hline
			WS 0.8 & 4      & 2      & 3      & 2      & \textbf{1}      & 0      & 0      & 0      & 0  \\ \hline
			WS 0.9 & 11     & 16     & 13     & 13     & \textbf{3}      & 3      & 5      & 1      & 2  \\ \hline
		\end{tabular}%
	}
	\caption{Classification of $\mathit{900}$ \emph{ER} networks with $\mathit{n=50}$ nodes; $\mathit{100}$ networks as $\mathit{p}$ varies ($\mathit{0.1, 0.2 \ldots 0.9}$).}
 \label{Table5}
\end{table}

\begin{table}[H]
	\resizebox{\textwidth}{!}{%
		\begin{tabular}{llllllllll}
			\hline
		n=75	&   ER 0.1 & ER 0.2 & ER 0.3 & ER 0.4 & ER 0.5 & ER 0.6 & ER 0.7 & ER 0.8 & ER 0.9  \\ \hline
			ER     & 50     & 50     & 50     & 50     & 50     & 50     & 50     & 50     & 50 \\ \hline
		\end{tabular}%
	}
	\caption{Classification of $\mathit{450}$ \emph{ER} networks with $\mathit{n=75}$ nodes; $\mathit{50}$ networks as $\mathit{p}$ varies ($\mathit{0.1, 0.2 \ldots 0.9}$).}
 \label{Table6}
\end{table}

\begin{table}[H]
	\resizebox{\textwidth}{!}{%
		\begin{tabular}{lllllllllll}
			\hline
		n=75	&    WS 0.6,10 & WS 0.6,11 & WS 0.6,12 & WS 0.6,13 & WS 0.6,14 & WS 0.6,15 & WS 0.6,16 & WS 0.6,17 & WS 0.6,18 & WS 0.6,19 \\ \hline
			ER        & 0         & 0         & 0         & 0         & 0         & 0         & 0         & 0         & 0         & 0  \\ \hline
			WS 0.0    & 0         & 0         & 0         & 0         & 0         & 0         & 0         & 0         & 0         & 0  \\ \hline
			WS 0.1    & 0         & 0         & 0         & 0         & 0         & 0         & 0         & 0         & 0         & 0  \\ \hline
			WS 0.2    & 0         & 0         & 0         & 0         & 0         & 0         & 0         & 0         & 0         & 0  \\ \hline
			WS 0.3    & 0         & 0         & 0         & 0         & 0         & 0         & 0         & 0         & 0         & 0  \\ \hline
			WS 0.4    & 0         & 0         & 0         & 0         & 0         & 0         & 0         & 0         & 0         & 0  \\ \hline
			WS 0.5    & 2         & 5         & 3         & 7         & 7         & 3         & 1         & 8         & 4         & 3  \\ \hline
			WS 0.6    & 40        & 38        & 44        & 37        & 34        & 43        & 45        & 35        & 40        & 41 \\ \hline
			WS 0.7    & 8         & 7         & 3         & 6         & 9         & 4         & 4         & 7         & 6         & 6  \\ \hline
			WS 0.8    & 0         & 0         & 0         & 0         & 0         & 0         & 0         & 0         & 0         & 0  \\ \hline
			WS 0.9    & 0         & 0         & 0         & 0         & 0         & 0         & 0         & 0         & 0         & 0  \\ \hline
		\end{tabular}%
	}
	\caption{Classification of $\mathit{500}$ \emph{WS} $\mathit{(0.6, k)}$ networks with $\mathit{n=75}$ nodes for different values of $\mathit{k}$.}
 \label{Table7}
\end{table}

\begin{table}[H]
	\resizebox{\textwidth}{!}{%
		\begin{tabular}{lllllllllll}
			\hline
		n=75	&    WS 0.9,10 & WS 0.9,11 & WS 0.9,12 & WS 0.9,13 & WS 0.9,14 & WS 0.9,15 & WS 0.9,16 & WS 0.9,17 & WS 0.9,18 & WS 0.9,19 \\ \hline
			ER        & \textbf{1}         & 0         & \textbf{1}         & 0         & 0         & \textbf{1}         & \textbf{1}         & \textbf{1}         & \textbf{1}         & \textbf{3}  \\ \hline
			WS 0.0    & 0         & 0         & 0         & 0         & 0         & 0         & 0         & 0         & 0         & 0  \\ \hline
			WS 0.1    & 0         & 0         & 0         & 0         & 0         & 0         & 0         & 0         & 0         & 0  \\ \hline
			WS 0.2    & 0         & 0         & 0         & 0         & 0         & 0         & 0         & 0         & 0         & 0  \\ \hline
			WS 0.3    & 0         & 0         & 0         & 0         & 0         & 0         & 0         & 0         & 0         & 0  \\ \hline
			WS 0.4    & 0         & 0         & 0         & 0         & 0         & 0         & 0         & 0         & 0         & 0  \\ \hline
			WS 0.5    & 0         & 0         & 0         & 0         & 0         & 0         & 0         & 0         & 0         & 0  \\ \hline
			WS 0.6    & 0         & 0         & 0         & 0         & 0         & 0         & 0         & 0         & 0         & 0  \\ \hline
			WS 0.7    & 0         & 2         & 0         & 0         & 0         & 0         & 0         & 0         & 0         & 0  \\ \hline
			WS 0.8    & 12        & 14        & 10        & 16        & 13        & 13        & 11        & 15        & 8         & 10 \\ \hline
			WS 0.9    & 37        & 34        & 39        & 34        & 37        & 36        & 38        & 34        & 41        & 37 \\ \hline
		\end{tabular}%
	}
	\caption{Classification of $\mathit{500}$ \emph{WS} $\mathit{(0.9, k)}$ networks with $\mathit{n=75}$ nodes for different values of $\mathit{k}$.}
 \label{Table8}
\end{table}

\paragraph{Interpretation}

The effectiveness of the classification method is more sound for larger networks, as evident in Tables~\ref{Table4},\ref{Table5}, and~\ref{Table6}. This phenomenon can be attributed to Theorem~\ref{thm:Rado}, which ensures the uniqueness of the Rado graph and the resilience of the neighborhood surrounding the critical point. Conversely, as observed in Tables~\ref{Table7} and~\ref{Table8}, the method demonstrates robustness in the opposite direction. The majority of networks are accurately mapped to the corresponding small-world model with the parameters used for their generation.

Let us consider the distinction between the pure random and the biased graph that we introduced in the abstract. In Table~\ref{Table5}, we see that, for $\mathit{n=50}$ and $\mathit{p=0.5}$, there still exist $\mathit{4}$ networks that look biased (numbers in bold). The latter leads us to the fact that any bias can be can be introduced arbitrarily. Conversely, in Table~\ref{Table8}, we see that biased networks can also look random (numbers in bold).


\section{Application to Temporal Networks}~\label{sec:app}

We apply RI to temporal networks; networks with a standard number of nodes and whose connectivity changes throughout time i.e. links between any pair of nodes can appear or disappear between two time steps. In Section~\ref{sec:Derivation} we discuss the process of deriving a temporal correlation network from a dynamical system studied by means of multivariate time series. Here, the connections between the nodes represent correlation. In Section~\ref{sec:stocknets}, we apply RI to stock networks~\cite{HUANG20092956}.

\subsection{Derivation of Temporal Correlation Network}~\label{sec:Derivation}

Each node $\mathit{v_i}$ of the derived temporal networks represents a vector $\mathit{x_i}$ of $\mathit{n}$ observed values which correspond to measurements of time series data:

\[
\Bar{x}_{i} = [x_{i}(1), \ldots x_{i}(n)]
\]

\begin{example}
    We consider a dataset that contains the Morgan Stanley Capital International's (MSCI) market capitalization weighted index of $\mathit{55}$ developed markets. This means that the derived network consists of $\mathit{55}$ nodes i.e. $\mathit{V=\{v_1, \ldots v_{55}\}}$, where each node $\mathit{v_i}$ corresponds to a vector of values $\mathit{\Bar{x}_i}$ of a particular market. Each vector $\mathit{\Bar{x}_i}$ comprises $\mathit{1305}$ daily indices for each market in the period 5 of March 2004 until 5 of March 2009, excluding weekends and holidays. Thus, $\Bar{x}_i = [x_{i}(1), \ldots, x_{i}(1305)]$. For example, at time $t=15$ the market $i$ has the value $x_{i}(15)$.
\end{example}

We first apply a procedure of standard preprocessing steps to the original time series data. The overall procedure is summarized in Fig.~\ref{fig:preprocessing}.

\begin{figure}[h]
    \centering
    \begin{equation*}
    \Bar{x}_{i}  \xrightarrow{\text{Dismiss Trends and Periodicity}} \Bar{y}_{i}  \xrightarrow{\text{Prewhitening}} \Bar{z}_{i}
    \end{equation*}
    \caption{Dismiss Trends and Periodicity from the original timeseries (first arrow), and then prewhiten (second arrow).}
    \label{fig:preprocessing}
\end{figure}

\paragraph{Dismiss Trends and Periodicity} 
In order to relieve the time series from tendency and periodicity, we compute the first differences of the logarithms of the original time series for the shake of stationarity: 
\begin{align}~\label{eq:differences}
    y_i(t-1)=log\bigl(x_{i}(t)\bigr)-log\bigl(x_i(t-1)\bigr), \forall i \land t \in [2, \ldots, 1305]
\end{align}
The difference reduces the tendency and the use of the logarithm decreases the variance. Therefore, we obtain the relevant returns $\mathit{\Bar{y}_i}$ for every market $\mathit{i}$.

\paragraph{Timeseries Prewhitening}
The prewhitening process removes the autocorrelations of the timeseries $\mathit{\bar{y}_i}$ which may cause spurious cross-correlations. Firstly, the mean is subtracted from each $\mathit{\bar{y}_i(t)}$ using the formula $\bar{y}'_i = \bar{y}_i - \bar{\mu}_i$ where $\mathit{\bar{\mu}_i}$ represents a vector whose entries is the mean value of $\mathit{\bar{y}_i}$. Subsequently, an autoregressive (AR) model of order $\mathit{p=20}$ is fitted to each $\mathit{\bar{y}'_i}$ time series using the \emph{arx()} function in Matlab. The AR model provides coefficients for the estimated model. Predicted values $\mathit{\bar{y}''_i}$ are obtained using the \emph{predict()} function, and the mean subtracted earlier is added back to yield $\mathit{\bar{y}'''_i}$. 
Finally, the residuals $\mathit{\bar{z}_i}$ are calculated by subtracting the predicted values from the observed values i.e. $\mathit{\bar{z}_i = \bar{y}'''_i - \bar{y}_i, \forall i}$. 
To conclude, the final result of the whole transformation is an array of residuals $\mathit{\bar{z}_i}$. Once the residuals are effectively whitened, they can be analyzed using standard statistical techniques without the complication of autocorrelation. We explain more analytically the prewhitening process in~\ref{sec:prewhitening}.

\paragraph{Dataset partition}
We now partition the residuals $\mathit{\Bar{z}_i}$ into an equal amount of chronologically consecutive measurements:
\begin{align*}
    \Bar{z}_{i_1} &= [z_{i}(1), \ldots,  z_{i} (k)]\\
    \Bar{z}_{i_2} &= [z_{i}(k+1), \ldots,  z_{i} (2k)]\\
    \ldots\\
    \Bar{z}_{i_m} &= [z_{i}\bigl((m-1) \cdot k+1\bigr), \ldots, z_{i} (n)]\\
\end{align*} 
The $\mathit{\Bar{z}_{i_l}}$ represent the time windows, and a network is derived for each of them.

\begin{example}
Consider the MSCI dataset. We first perform prewhitening as described previously to obtain the residuals $\mathit{\Bar{z}_i}$. We then partition $\mathit{\Bar{z}_i}$ into $\mathit{87}$, each one of them containing $\mathit{15}$ values $\mathit{z_i(t)}$ in a chronological order:    $\Bar{z}_{i_1} = [z_{i}(1), \ldots,  z_{i} (15)],
    \Bar{z}_{i_2} = [z_{i}(16), \ldots,  z_{i} (30)],
    \ldots,
    \Bar{z}_{i_{87}} = [z_{i}(86 \cdot 15 + 1), \ldots, z_{i} (1305)]$.
\end{example}

\paragraph{Create the adjacency matrices}
For each $\mathit{\Bar{z}_{i_l}}$ with $\mathit{1 \leq l \leq m}$, we derive a correlation network, denoted by $\mathit{A(l)}$. In this network, the markets are represented by nodes while statistical significant cross-correlations according to the last measurement of the time window $\mathit{\Bar{z}_{i_l}}$ between the markets are added as links. 
All previous measurements are utilized for assessing the statistical significance of the cross-correlation at the last time-point. 
For the evaluation of statistical significance, we conduct a significance test based on the randomization of the original time series. 
This testing procedure involves randomizing the original time series and requires specifying a significance level 
$\mathit{\alpha}$, which determines the density of the correlation network. A larger 
$\mathit{\alpha}$ implies more statistically significant correlations and, consequently, higher density of the network.
The overall process of deriving the adjacency matrices is summarized in Fig.~\ref{fig:adj}.

\begin{figure}[h]
    \centering
    \begin{equation*}
    \Bar{z}_{i}  \xrightarrow{\text{Paritioning}} [\Bar{z}_{1}, \ldots, \Bar{z}_{m}]  \xrightarrow{\text{Correlation Matrices}} [A(1), \ldots, A(m)]
\end{equation*}
    \caption{To derive the adjacency matrices, we first (left arrow) partition the residuals according to time windows, and then (right arrow) compute the adjacency matrices.}
    \label{fig:adj}
\end{figure}


\subsection{Application to Stock Networks}~\label{sec:stocknets}

\paragraph{Hypothesis and Dataset}
We conduct an econometric application by considering stock networks; networks whose nodes correspond to stock indexes while edges correspond to statistically significant correlation between the stock indexes. We propose that our framework can detect anomalies and change-points in systems studied by means of multivariate time series.


\paragraph{Configuration}
We partition the $\mathit{55}$ time-sequence vectors of the MSCI dataset into $\mathit{29}$ and $\mathit{87}$ subvectors, each one of them containing $\mathit{45}$ and $\mathit{15}$ returns in a time row. That is:
\begin{align*}
    \Bar{z}_{i_1} &= [z_{i}(1), \ldots,  z_{i} (45)]\\
    \Bar{z}_{i_2} &= [z_{i}(46), \ldots,  z_{i} (90)]\\
    \ldots\\
    \Bar{z}_{i_{29}} &= [z_{i}(28 \cdot 45 + 1), \ldots, z_{i} (1305)]\\
\end{align*}
while in the second case we have:
\begin{align*}
    \Bar{z}_{i_1} &= [z_{i}(1), \ldots,  z_{i} (15)]\\
    \Bar{z}_{i_2} &= [z_{i}(16), \ldots,  z_{i} (30)]\\
    \ldots\\
    \Bar{z}_{i_{87}} &= [z_{i}(86 \cdot 15 + 1), \ldots, z_{i} (1305)]\\
\end{align*}

\begin{figure}[h]
    \centering
    \begin{tikzpicture}
        \node (element1) at (0,0) {$\mathit{A(l)}$};
        \node (element2) at (4,2) {$\mathit{\calF_{A(l)}}$};
        \node (element3) at (4,-2) {$\mathit{\calF_{\calG}(55,d_{A(l)})}$};
        \node (element4) at (8,0) {$\mathit{\norm{\calF_{A(l)}-\calF_{\calG}(55,d_{A(l)})}}$};
        
        \draw[->] (element1) -- node[above left] {} (element2);
        \draw[->] (element1) -- node[below left] {} (element3);
        \draw[->] (element2) -- node[above right] {} (element4);
        \draw[->] (element3) -- node[below right] {} (element4);
    \end{tikzpicture}
    \caption{The computation of randomness at one glance.}
    \label{fig:rhombus-diagram}
\end{figure}
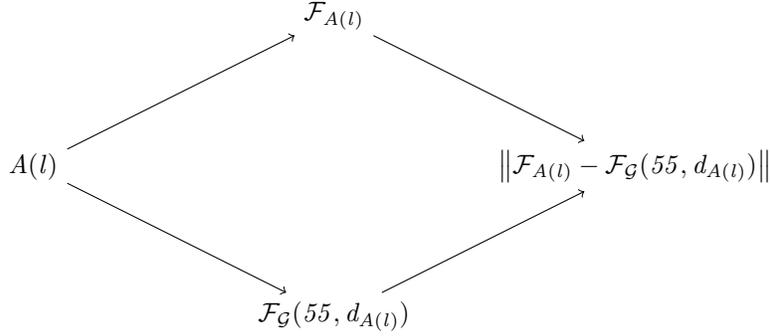

\paragraph{Compute randomness} 
In order to compute the randomness of each element we follow Fig.~\ref{fig:rhombus-diagram}.
For each $\mathit{A(l)}$, we compute the relevant RFP denoted by $\mathit{\calF_{A(l)}}$ (as described in Routine~\ref{routine: ComputeRFP} and displayed in the upper left arrow of Fig.~\ref{fig:rhombus-diagram}). We also compute the density of $\mathit{A(l)}$ denoted by $\mathit{d_{A(l)}}$ and insert it as the probability parameter for the calculation of the critical point $\mathit{\calF_{\calG}(55,d_{A(l)})}$ (see bottom left arrow of Fig.~\ref{fig:rhombus-diagram}). The randomness for a particular time window $\mathit{l}$ is the Euclidean distance:
\begin{align}
\norm{\calF_{A(l)}-\calF_{\calG}(55,d_{A(l)})}
\end{align}

\begin{figure}[H]
\centering
\resizebox{\textwidth}{!}{\includegraphics{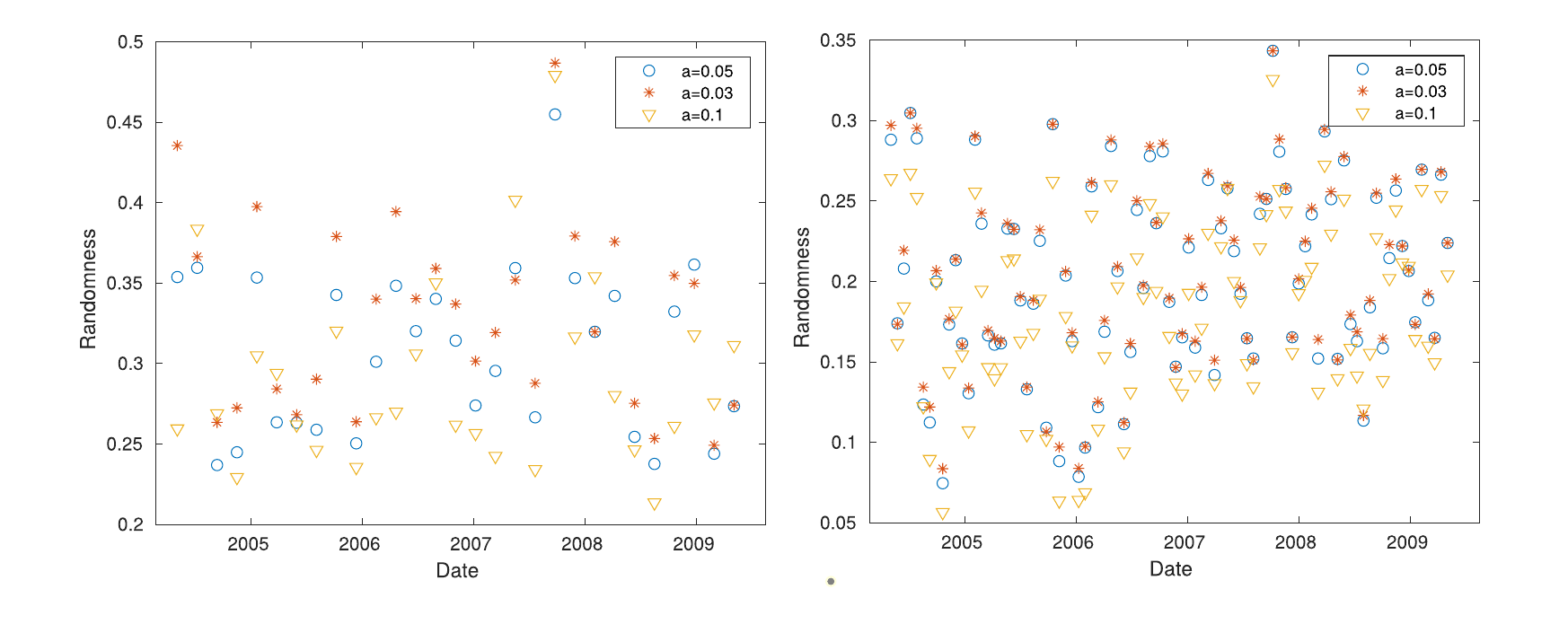}}
\caption{(Left) The values of $\norm{\calF_{A(l)}-\calF_{\calG}(55,d_{A(l)})}$ for $\mathit{29}$ snapshots. (Right)  The values of $\norm{\calF_{A(l)}-\calF_{\calG}(55,d_{A(l)})}$ for $\mathit{87}$ snapshots.}
\label{fig:res}
\end{figure}

\newpage
\paragraph{Results} 
We summarize the results in Fig.~\ref{fig:res}. 

The x-axis are the dates and the y-axis the value of randomness.
The figures display the results for different values of $\mathit{\alpha}$ as we display in the legends in the upper-right part of each figure. Particularly, the empty circles are the values of randomness in the case that $\mathit{\alpha=0.05}$, the full circle in the case that $\mathit{\alpha=0.03}$, and the triangle in the case that $\mathit{\alpha=0.1}$.

\paragraph{Interpretation} As we see in Fig.~\ref{fig:res}, the peak is reached in August 2007 (just before 2008 in the X-axis), marking the onset of the 2007-2008 global financial crisis-a profound economic downturn that unfolded in the early 21st century. It's worth noting that the zenith remains consistent across the three selected values of the parameter $\mathit{\alpha}$. Even when the dataset is divided into 29 subsets instead of 87, the results regarding the timing of the NRI's maximum value are unchanged. 
The application to stock networks, in this case, shows that NRI is high during recession periods; fact that reassures about the robustness of NRI. We propose that NRI can be utilized for anomaly detection in dynamical systems studied by means of multivariate time series.

\section{Conclusion and Future Work}~\label{sec: conclusion}

We propose a graph embedding technique where we represent a graph as a point in the $\mathit{6}$ dimensional space. Each coordinate in this space is the relative frequency of a particular network motif. By defining the unique point of the random (Rado) graph in the same $\mathit{6D}$ space, we can quantify the randomness of an observed network the proximity of its point to the point of the random graph.
Motifs drive network randomness beyond probability. 

\paragraph{Future Work} By extending to directed graphs, we can study the randomness of causality networks. It will be also interesting to apply the RI to other dynamical systems studied under means of multivariate time series (like EEGs).
Some rough simulations that we performed, show that multivariate time series with trend and autocorrelation construct random networks in contrast with prewhitened time series (noisy signals) that tend to construct ordinary lattices. Instead of the Pearson correlation coefficient and the prewhitening, one could also use the mutual information. Last but not least,
an extension to graphlets of size $\mathit{5}$ is also feasible.

\paragraph{Related Work} The initial work was containing comparative analysis and separation between three different network types: random, small-world and scale-free networks. For the case of scale-free networks we considered two different generators; the Barabasi-Albert mechanism and a mechanism that was generating scale-free networks with a power-law degree distribution of a user-defined exponent. However, scale-free networks can be recognised to the naked eye, so we dismiss this analysis. For a motif analysis of different generators of scale-free networks, we refer the reader to~\cite{zhang2015exactly}. Moreover in~\cite{zhang2015exactly,judd2013exactly,small2014complex}, the authors experimentally prove that scale-free networks display a wider spectra of properties that the Barabasi-Albert preferential attachment mechanism fails to reproduce.
Recent evidence propose that the node degree distribution of scale-free networks is heavy-tailed and not always power-law.

\section*{Acknowledgements}

Part of this work has been done during the master on Web Science at the School of Sciences, Department of Mathematics of Aristotle University of Thessaloniki in Greece (2014-2016, see thesis~\citep{Argyris:287148}). Thanks to Dimitris Kugiumtzis for his guidance in the very early steps of this work, for providing some pieces of matlab codes and for providing the Morgan Stanley Capital Investment dataset. 
The work was also partially supported by the Ph.D. school of education of the Technical University of Denmark (Department of Applied Mathematics and Computer Science, Section Software Systems Engineering) while the writer of current manuscript was a Ph.D. student.

\bibliography{main}

\begin{appendices}
~\label{Appendices}

\section{Empirical Construction of $\calF_{\calW}$}~\label{appendix:A}

The algorithm for the empirical construction of $\calF_{\calW}$ incorporates two routines/functions: the \emph{CreateSmallWorldGraph} routine and the  \emph{ComputeRFP} routine. All implementations can be found on~\url{https://github.com/GeorgiosArg/Diagnosis-of-Small-World-Bias-in-Random-Graphs}.

\subsection{CreateSmallWorldGraph Routine}~\label{routine: CreateSmallWorldGraph}The \emph{CreateSmallWorldGraph} routine is a function taken from from~\cite{muchnik2013complex}. It inputs the Small-world parameters $(n,k,p_r)$ of Def.~\ref{def:swgraph} and generates a small-world graph according to the Watts-Strogatz model.

\subsection{ComputeRFP Routine}~\label{routine: ComputeRFP} For the shake of this work, we implemented the \emph{ComputeRFP} routine which inputs a network and outputs its RFP. This routine first enumerates the occurrences of each one of the motifs of Fig.~\ref{fig:motifs}. The initial implementation inputs the graph as a matrix and iterates over all different tetrads (i.e. 4-tuples) of nodes, i.e. each $\mathit{4\times4}$ submatrix of the matrix represantion of the network, and maps it to the corresponding motif of Fig.~\ref{fig:motifs}-if any of these are constructed-. This is computationally expensive. 
Each motif of Fig.~\ref{fig:motifs} can be represented by more than one adjacency matrix; and the adjacency matrices are as many as the different patterns of each motif in Fig.~\ref{fig:patterns}. To overcome this problem, smart implementations like the Brain Connectivity Toolbox~\cite{RUBINOV20101059} use dictionaries as the data structure to represent graphs. We incorporated this implementation and in order to further reduce complexity, we employed one more fact: \emph{For graphs with equal size $n$, and $n\leq 4$ it holds that: the graphs have the same degree distribution if and only if they are isomorphic} (see Fig.~\ref{fig:patterns}).

\subsection{The Construction Algorithm}

\begin{algorithm}[htp]
	\SetAlgoLined
	\KwResult{A list containing $F_{\calW}$ for different values of $n, k, p_r$}
	Define network size $n$;\\
	$m \leftarrow 100$;\\
	\For{$p_r$ from $0$ to $0.9$ with step $0.1$}{
		\For{$k$ from $2$ to $k_{max}$ with step $2$}{
			$sum = (0, 0, 0, 0, 0, 0)$;\\
			\For{b=1:m}{
				$Graph$ = CreateSmallWorldGraph($n$,$k$,$p_r$)\\
				$\calF_{o}$ = ComputeRFP($Graph$)\\
				$sum=sum+\calF_{o}$\\}
			$\calF_{\calW}(n,k,p_r)=sum/m$;\\
			$list_{\calF_{\calW}}=list_{\calF_{\calW}}.append(\calF_{\calW}(n,k,p_r))$\\
		}
	}
	\caption{Computation of the RFP $\calF_{\calW}(n,k,p_r)$.}
	\label{alg:RFP_SW}
\end{algorithm}

The $\mathit{m}$ determines how many networks we generate or, in other words, the number of RFPs that we average.
When we explore networks of size $\mathit{n=25}$, and $\mathit{n=50}$, we generate $\mathit{100}$ networks, while for size $\mathit{n=75}$ we average over $\mathit{50}$ networks due to the complexity of the \emph{ComputeRFP} routine that we explain later.

Inside the two \emph{for-loops} which iterate over $\mathit{p_r}$ and $\mathit{k}$, we compute the RFP for that specific $\mathit{p_r}$ and $\mathit{k}$. $\mathit{p_r}$ iterates over $\mathit{10}$ possible values, i.e., $\mathit{0, 0.1, \ldots, 0.9}$. Thereafter, the algorithm iterate over $\mathit{k}$ from $\mathit{2}$ to $k_{max}$.
The $k_{max}$ in the second iteration of this algorithm refers to the \emph{possible values} that the parameter $\mathit{k}$ can obtain which increases with respect to $\mathit{n}$; the more the nodes of the circle, the more potential neighbors a node can obtain. The step is 2 because $\mathit{k}$ has to be an even number since every node is connected to the two nearest neighbours (one to the left and one to the right) in the circular configuration of the Watts-Strogatz model. It has to hold that $\mathit{k_{max}<n}$, i.e., the degree of a node $k_{max}$ must not exceed the size of the network. Hence, $\mathit{k_{max}}$ is given by the following piecewise function:

\[ 
k_{max} = \begin{cases}
n-2 & \mbox{ if $mod(n,2)=0$}\\
n-1 & \mbox{ if $mod(n,2)=1$}
\end{cases}
\]

So sum up, the length of the $list_{\calF_{\calW}}$ (the amount of the generated RFPs) depends on the size ($\mathit{n=|V|}$) of the obseved network that we aim to classify. 
For example, if the observed that we want to classify has size $\mathit{55}$, the value of $\mathit{k_{max}}$ is $\mathit{54/2}$.  Hence, the list contains $\mathit{10 \times 27=270}$ RFPs because $\mathit{p_r}$ can take $\mathit{10}$ different values $\mathit{0, 0.1, \ldots, 0.9}$.

\section{Construction of the list of the RFPs that correspond to the ER and SW network topologies}~\label{sec:RFP_top}

\begin{algorithm}[htp]
	\SetAlgoLined
	\KwResult{A list containing the RFPs of the network topologies}
	Define network size $n$;\\
	\For{$p$ from $0$ to $0.9$ with step $0.1$}{
		Compute $\calF_{\calG}(n,p)$ (Section~\ref{subsec:CP}, Lemma~\ref{lem:RF});\\ 
		$list=list.append(\calF_{\calG}(n,p))$;\\
		}
	$m \leftarrow 100$;\\
	\For{$p_r$ from $0$ to $0.9$ with step $0.1$}{
		\For{$k$ from $2$ to $max\_k$ with step $2$}{
			$sum = (0, 0, 0, 0, 0, 0)$;\\
			\For{b=1:m}{
				$Graph$ = CreateSmallWorldGraph($n$,$k$,$p_r$)\\
				$\calF_{o}$ = ComputeRFP($Graph$)\\
				$sum=sum+\calF_{o}$\\}
			$\calF_{\calW}(n,k,p_r)=sum/m$;\\
			$list=list.append(\calF_{\calW}(n,k,p_r))$\\
		}
	}
	\caption{The embedding Algorithm; Computation of the RFPs of the network topologies.}
	\label{alg:Embeding}
\end{algorithm}

For example, if $n=25$ we constructed $\mathit{9}$ RFPs for the ER graph for different values of $\mathit{0.1 \ldots 0.9}$ and $number$ of RFPs for the WS graph for different values of $\mathit{p_r}$, and $\mathit{k}$. The list in this case contains $number$ RFPs.
Similarly, for $n=50$ the list contains $number$ and for $n=75$ the list contains $number$ RFPs.
The routines \emph{CreateSmallWorldGraph}, and \emph{ComputeRFP} are explained in the previous Appendix~\ref{appendix:A}, in the corresponding Sections~\ref{routine: CreateSmallWorldGraph},~\ref{routine: ComputeRFP}.
All implementations can be found on~\url{https://github.com/GeorgiosArg/Diagnosis-of-Small-World-Bias-in-Random-Graphs}.

\section{The Classification Algorithm}~\label{appendix:B}

\begin{algorithm}[htp]
	\SetAlgoLined
	\KwResult{The generative model of a given graph.}
	Import an observed network $G=(V,E)$;\\
	$\calF_G$ = Compute$RFP$($G$); \# the routine of Section~\ref{routine: ComputeRFP}. \\
	$\calF_{\calW}$ = Compute$RFP_{\calW}$ ($|V|$); \# the empirical construction of~\ref{appendix:A}.\\
	$\calF_{\calG}$ = Compute$RFP_{\calG}$ ($|V|$); \# Calculated according to Section~\ref{subsec:CP}, Lemma~\ref{lem:RF}.\\ 
	\eIf{$|\calF_G-\calF_{\calG}|$ $<$  $|\calF_G-\calF_{\calW}|$}{ ``Erdos-Renyi is the generative model"}{``Watts-Strogatz is the generative model"}
	\caption{Guessing the generative model.}
	\label{alg:classification_method}
\end{algorithm}

\section{Further Analysis of the Classification Method}

\begin{enumerate}[Step 1:]
	\item Construction of the RFP for each model and for each one of their parameters so as to compute the corresponding averaged point $\calF_j$:
	\begin{enumerate}
		\item $j = 1, \ldots,9$ for $ER(p)$ when $p=0.1, 0.2, \ldots, 0.9$
		\item $j = 10, 11, \ldots, \floor{\frac{n}{2}} \times 10 $ for $WS$ when $k=2, 4, 6, \ldots, \floor{\frac{n}{2}} $ \newline and $ p= 0, 0.1,
		0.2, \ldots, 0.9 $.
	\end{enumerate}
	\item Computing the corresponding RFP of the observed network $G$ which is denoted as $\calF_G$.
	\item The generative model/target function is the one that minimizes the euclidean distance between point $\calF_G$ and the point constructed in step 1:\begin{align*}
	\minimize_j \norm{\calF_G-\calF_j}
	\end{align*}
\end{enumerate}

In other words, we first embed the topologies into their \emph{relative frequency points (RFPs)} which are 6-dimensional points with each dimension corresponding to the relative frequency of a motif. The topology of an observed network is determined by the following steps: (i) compute the RFP of the observed network, (ii) compute the Euclidean distance between the RFP of the observed network and the RFPs that characterize the different network topologies, and (iii) the topology is the one that its RFP minimizes the Euclidean distance with the RFP of the observed network.

\section{Prewhitening}~\label{sec:prewhitening}

The prewhitening removes the autocorrelations of the timeseries $\mathit{\bar{y}_i}$ which may cause spurious cross-correlations. We first subtract the mean as follows:
\begin{align}~\label{eq:mean}
    \bar{y}'_i = \bar{y}_i - \bar{\mu}_i, \forall i
\end{align}
where $\mathit{\bar{\mu}_i}$ is an array whose entries $\mathit{\bar{\mu}_i(t), t \in \{1, \ldots, 1305\}}$ is the mean $\mathit{\mu_i}$ of $\mathit{\bar{y}_i}$ averaged over all $\mathit{t}$. We now fit an autoregressive (AR) model to the timeseries $\mathit{\bar{y}_i}$. We selected an autoregressive model of order $\mathit{p=20}$ by default.
\begin{align}~\label{eq:ARmodel}
    AR = arx(\bar{y}'_i,p), \forall i
\end{align}
The \emph{arx()} is a built-in function of Matlab incorporated into the System Identification Toolbox which fits an autoregressive model of order $\mathit{p}$ to the time series $\mathit{\bar{y}'_i 
= \bar{y}_i - \bar{\mu}_i}$. The resulting \emph{AR} object contains the coefficients of the estimated model.
By feeding the AR model with the timeseries $\mathit{\bar{y}'_i}$ , we obtain the predicted values $\mathit{\bar{y}''_i}$:
\begin{align}~\label{eq:predict}
    \bar{y}''_i = predict(AR,\bar{y}'_i), \forall i
\end{align}
The \emph{predict()} is another built-in function of Matlab incorporated into the System Identification Toolbox. \emph{AR} is the autoregressive model obtained from the \emph{arx()} function of equation~(5), $\mathit{\bar{y}'_i}$ The input data used for prediction.
We then add the mean that we subtracted in equation~(4).
\begin{align}~\label{eq:predicted_values}
    \bar{y}'''_i = \bar{y}''_i+\bar{\mu}_i, \forall i
\end{align}
Finally, we proceed with the calculation of the residuals; the part of the original series that is not explained by the fitted model. We calculate the residuals by subtracting the predicted values from the observed values.
\begin{align}~\label{eq:residuals}
    \bar{z}_i = \bar{y}'''_i - \bar{y}_i, \forall i
\end{align}
The final result of the whole transformation is an array of residuals $\mathit{\bar{z}_i}$. Once the residuals are effectively whitened, they can be analyzed using standard statistical techniques without the complication of autocorrelation. All implementations can be found on~\url{https://github.com/GeorgiosArg/Diagnosis-of-Small-World-Bias-in-Random-Graphs}.
 
\end{appendices}

\end{document}